\documentclass[aps,pra,superscriptaddress,amsmath,amssymb,reprint,floatfix]{revtex4-2}

\usepackage{tikz}
\usepackage{graphicx}
\usepackage{dcolumn}
\usepackage{bm}
\usepackage{float}
\usepackage{svg}
\usepackage{braket}
\usepackage{bbold}
\usepackage{enumerate}
\usepackage{wrapfig}
\usepackage{amsmath} 
\usepackage{amssymb} 
\usepackage{seqsplit}
\usepackage{xcolor}
\usepackage{braket}
\usepackage{tabularx}
\usepackage{makecell}
\usepackage[export]{adjustbox}
\usepackage[shortlabels]{enumitem}
\usepackage{setspace}
\usepackage{tensor}
\usepackage{subfigure}
\usepackage{hyperref}
\hypersetup{
    colorlinks=true,       
    linkcolor=cyan,          
    citecolor=magenta,        
    filecolor=magenta,      
    urlcolor=cyan,           
    runcolor=cyan
}
\usepackage[capitalise]{cleveref} 
\usepackage{chngcntr}

\newcounter{myfigure}
\counterwithin*{figure}{myfigure}

\newcommand{\bes} {\begin{subequations}}
\newcommand{\ees} {\end{subequations}}
\def\Tr{\mathrm{Tr}}

\def\theYear{\the\year}
\hypersetup{pdfpagemode=FullScreen, colorlinks=true, allcolors=blue}

\begin{document}

\title{Beating the Ramsey limit on sensing with deterministic qubit control}
\author{Malida Hecht}
\affiliation{Center for Quantum Information Science and Technology, University of Southern California, Los Angeles, California 90089}
\affiliation{Department of Physics \& Astronomy, University of Southern California, Los Angeles, California 90089}

\author{Kumar Saurav}
\affiliation{Center for Quantum Information Science and Technology, University of Southern California, Los Angeles, California 90089}
\affiliation{Ming Hsieh Department of Electrical \& Computer Engineering, University of Southern California, Los Angeles, California 90089}

\author{Evangelos Vlachos}
\affiliation{Center for Quantum Information Science and Technology, University of Southern California, Los Angeles, California 90089}
\affiliation{Department of Physics \& Astronomy, University of Southern California, Los Angeles, California 90089}

\author{Daniel A. Lidar}
\affiliation{Center for Quantum Information Science and Technology, University of Southern California, Los Angeles, California 90089}
\affiliation{Department of Physics \& Astronomy, University of Southern California, Los Angeles, California 90089}
\affiliation{Ming Hsieh Department of Electrical \& Computer Engineering, University of Southern California, Los Angeles, California 90089}
\affiliation{Department of Chemistry, University of Southern California, Los Angeles, California 90089}
\author{Eli M. Levenson-Falk}
\email{elevenso@usc.edu}
\affiliation{Center for Quantum Information Science and Technology, University of Southern California, Los Angeles, California 90089}
\affiliation{Department of Physics \& Astronomy, University of Southern California, Los Angeles, California 90089}
\affiliation{Ming Hsieh Department of Electrical \& Computer Engineering, University of Southern California, Los Angeles, California 90089}

\begin{abstract}
Quantum sensors promise revolutionary advances in medical imaging \cite{aslamQuantumSensorsBiomedical2023}, energy production \cite{crawfordQuantumSensingEnergy2021}, mass detection \cite{asenbaumPhaseShiftAtom2017}, geodesy \cite{grottiGeodesyMetrologyTransportable2018},  foundational physics research \cite{dixitSearchingDarkMatter2021,bothwellResolvingGravitationalRedshift2022,bassQuantumSensingParticle2024}, and a host of other fields. In many sensors, the signal takes the form of a changing qubit frequency, which is detected with an interference measurement. Unfortunately, environmental noise decoheres the qubit state, reducing signal-to-noise ratio (SNR). Here we introduce a protocol for enhancing the sensitivity of a measurement of a qubit's frequency in the presence of decoherence. We use a continuous drive to stabilize one component of the qubit's Bloch vector, enhancing the effect of a small static frequency shift. We demonstrate our protocol on a superconducting qubit, enhancing SNR per measurement shot by $1.65\times$ and SNR per qubit evolution time by $1.09\times$ compared to standard Ramsey interferometry. We explore the protocol theoretically and numerically, finding maximum enhancements of 1.96$\times$ and 1.18$\times$, respectively. We also show that the protocol is robust to parameter miscalibrations. Our protocol provides an unconditional enhancement in signal-to-noise ratio compared to standard Ramsey interferometry. It requires no feedback and no extra control or measurement resources, and can be immediately applied in a wide variety of quantum computing and quantum sensor technologies to enhance their sensitivities.
\end{abstract}

\maketitle
\section{Introduction}\label{sec:intro}
Ramsey interferometry \cite{ramseyMolecularBeamResonance1950} has been long established as the most sensitive measure of a qubit's frequency \cite{degenQuantumSensing2017}. In a Ramsey measurement, a qubit is prepared in a superposition of energy states, allowed to evolve freely and acquire phase, and then measured along some axis. The phase acquired (i.e., the measured state probability) depends on the qubit frequency. This protocol has been used for quantum sensing of magnetic field and other continuous variables \cite{budkerOpticalMagnetometry2007,balasubramanianNanoscaleImagingMagnetometry2008,balUltrasensitiveMagneticField2012}, for foundational physics \cite{dixitSearchingDarkMatter2021,bassQuantumSensingParticle2024}, for biomedical applications \cite{aslamQuantumSensorsBiomedical2023}, for detection of nonequilibrium quasiparticle densities \cite{risteMillisecondChargeparityFluctuations2013,serniakHotNonequilibriumQuasiparticles2018a,liuQuasiparticlePoisoningSuperconducting2024}, and for rapid recalibration of a qubit's frequency \cite{vepsalainenImprovingQubitCoherence2022}, among many other applications. Decoherence at a rate $\gamma_2 = 1/T_2$ limits the signal-to-noise ratio (SNR) of quantum sensors. To date, most work has focused on SNR \emph{scaling} beyond the standard quantum limit of $\sqrt{N}$ (where $N$ is the number of independent qubits and/or measurements) \cite{jonesMagneticFieldSensing2009, tanakaProposedRobustEntanglementBased2015,matsuzakiMagneticFieldSensing2011,lawrieQuantumSensingSqueezed2019, zhangDistributedQuantumSensing2021}, using dynamical decoupling to enhance frequency discrimination and reduce non-Markovian decoherence \cite{bossQuantumSensingArbitrary2017, titumOptimalControlQuantum2021a}, using weak measurement feedback to rapidly lock in on time-varying signals \cite{naghilooAchievingOptimalQuantum2017a} or an unknown signal axis \cite{songAgnosticPhaseEstimation2024}, and compensating for measurement errors \cite{lenQuantumMetrologyImperfect2022}. Likewise, researchers have developed sensors which are less prone to decoherence \cite{shulmanSuppressingQubitDephasing2014,pengReductionSurfaceSpininduced2020,danilinQuantumSensingTunable2024a} and couple more strongly to wanted signals \cite{taylorHighsensitivityDiamondMagnetometer2008,wernsdorferMicroNanoSQUIDsApplications2009,levenson-falkDispersiveNanoSQUIDMagnetometry2016}. However, no results have shown improvement over traditional Ramsey interferometry for increasing the SNR from a \emph{single} qubit measuring a \emph{static} (zero frequency) signal.

Here, we demonstrate a protocol for enhanced quantum sensing of static fields. The protocol is based on a recent theory of quantum property preservation generalizing the coherence preservation results of \cite{LidarSchneider:04}, showing how certain scalar functions of a quantum state can be stabilized using purely Hamiltonian control~\cite{QPP}. We use deterministic Hamiltonian control of a single qubit to stabilize one Bloch vector component ($v_x)$, enabling increased phase accumulation in the orthogonal component $v_y$ and thus enhanced sensitivity. Our protocol gives a significant signal enhancement over standard Ramsey interferometry, up to a factor of 1.96 per measurement shot or 1.18 per qubit evolution time. We derive analytical expressions for the signal enhancement in the small-signal regime, and show simulations of the protocol's robustness to miscalibrations. The protocol is robust to variation in environmental parameters, requires no feedback (i.e., is unconditional and deterministic), and can be applied in a wide variety of experimental systems. Our results demonstrate a general technique for enhanced quantum sensing.

\begin{figure}
    \centering
    \includegraphics[width=3.3in]{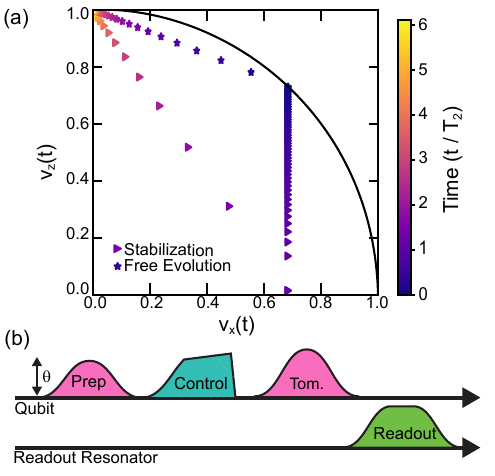}
    \caption{\label{fig:cartoons}
    (a) Simulated trajectories of the Bloch vector in the $xz$ plane for a qubit state which is either allowed to freely evolve (star markers) or stabilized with our protocol (triangle markers), with $T_1 = T_2 = 1$ and $v_{x}(0) = 0.68$. The black curve is the $|v|=1$ limit for a pure state. (b) Pulse sequence schematic of our stabilization protocol. We prepare a state in the $xz$ plane at an angle $\theta$ from the $z$ axis,  
    then stabilize the Bloch component $v_x = \sin\theta$ up to a breakdown time. After the breakdown we turn off the control and the state is allowed to freely decay. After a variable evolution time we perform quantum state tomography. 
    }
\end{figure}

\section{Sensing with Ramsey Interferometry}
Consider a sensor that linearly maps some environmental variable $B$ to a qubit's frequency, and thus to the detuning $\Delta$ between qubit frequency and drive. We assume the transduction function $\Delta = f(B)$ is a fixed and known property of the sensor. The qubit's state is given by the Bloch vector $\vec{v} = (v_x,v_y,v_z) = (\left<\sigma_x\right>,\left<\sigma_y\right>,\left<\sigma_z\right>)$. We choose coordinates such that the initial state lies in the $xz$ plane, $\vec{v}(0) = (v_x(0),0,v_z(0))$. Detuning causes rotation about the $z$ axis, leading $v_y$ to grow proportionally to $v_x$ at a rate $\Delta$. For small rotation angle, $v_y(t)$ is linearly proportional to $\Delta$---this is the relevant limit for a weak signal with $\Delta \ll \gamma_2 = 1/T_2$. Measuring $v_y$ is thus a measurement of $\Delta$. The measurement SNR for small $v_y$ is $\sqrt{N}v_y$, where $N$ is the number of measurement iterations. We assume that state preparation and measurement errors affect all protocols equally and, therefore, ignore them for this analysis.

The task is thus to maximize $\sqrt{N} v_y= \sqrt{\frac{T}{t+t_i}} v_y(t)$, where $T$ is the total experiment time, $t$ is the time the qubit spends accumulating phase in an iteration, and $t_i$ is the ``inactive'' time spent preparing, reading out, and resetting the qubit in an iteration. When $t_i \gg t \sim T_2$, $N$ is essentially fixed and the goal is to maximize $v_y(t)$; when $t \sim T_2 \gg t_i$, the goal is to maximize $v_y(t)/\sqrt{t}$.

Consider a qubit subject to relaxation at rate $\gamma_1 = 1 / T_1$, dephasing at rate $\gamma_\phi$, detuning $\Delta$, and a coherent drive $H_\mathrm{drive}(t) = h_y(t) \sigma_y$. In a Ramsey sequence the qubit is prepared in the state $\vec{v}(0) = (1,0,0)$ and allowed to freely evolve ($h_y = 0$). When $\Delta \ll \gamma_2 = \gamma_\phi + \gamma_1/2$, this leads to a maximum $y$-component 
\begin{equation}
\label{eq:ramsey-signal}
    v_y^{R,\max} \approx \frac{1}{e}\frac{\Delta}{\gamma_2}
\end{equation}
at time $t \approx T_2$, and a maximum $y$-component per root evolution time
\begin{equation}
\label{eq:ramsey-signal2}
\frac{v_y^R(t_{{\max}})}{\sqrt{t_{\max}}} = \frac{1}{\sqrt{2 e}}\frac{\Delta}{\sqrt{\gamma_2}} \approx 0.429 \frac{\Delta}{\sqrt{\gamma_2}}
\end{equation}
at $t\approx T_2/2$ (see \cref{app:stable} for derivations). We define the \emph{signal improvement ratio} $R_v$ (i.e., the SNR improvement per measurement shot) as the ratio of maximum $v_y$ from the given protocol to that from the Ramsey experiment, the latter being given by \cref{eq:ramsey-signal}. Likewise, we define the \emph{signal per root evolution time improvement ratio} $R_s$ (i.e., the SNR improvement per evolution time) as the ratio of maximum $v_y/\sqrt{t}$ from the given protocol to that from the Ramsey experiment, the latter being given by \cref{eq:ramsey-signal2}.

\section{Sensing with Coherence Stabilization}
Our protocol uses Hamiltonian control to preserve state coherence~\cite{LidarSchneider:04} and enhance sensitivity. The general theory is derived in \cite{QPP}. Here we report the central results, with full details given in \cref{app:stable,app:unstable}. Again, we have initial state $\vec{v}(0) = (v_x(0),0,v_z(0))$. For small, unknown $\Delta \ll \gamma_2$, we can stabilize $v_x(t) \approx v_x(0)$ for $0\le t\le t_b$ so long as $v_z(t) \neq 0$ by setting 
\begin{equation}\label{eq:control}
h_y(t) = \frac{\gamma_2 v_x(0)}{2 v_z(t)}
\end{equation}
where the stabilization is exact to 2nd order in $\Delta/\gamma_2$. In general, coherence will be transferred from $v_z$ to $v_x$ until $v_z\rightarrow 0$. At this \emph{breakdown time} $t_b$, the protocol fails and coherence must decay. However, stability can be achieved indefinitely if $v_x(0) \leq \frac{1}{2}\sqrt{\frac{\gamma_1}{\gamma_2}}$, since at low temperature relaxation deterministically transfers population to the ground state. The drive then transfers this population back to the desired $v_x$, preserving coherence. See \cref{fig:cartoons} for an illustration of a Bloch trajectory with coherence stabilization. Temperature effects are discussed in \cref{app:bloch,app:stable}.

When $v_x$ is thus coherence-stabilized at $v_x^c \approx v_x(0)$ and the unknown detuning $\Delta \neq 0$, $v_y$ grows to an asymptotic maximum $v_y^c \rightarrow v_x(0)\Delta/\gamma_2$, leading to a signal improvement ratio (SNR per measurement shot improvement ratio) relative to Ramsey of $R_v = e v_x(0)$. The maximum stable $v_x(0) = \frac{1}{2}\sqrt{\frac{\gamma_1}{\gamma_2}}$ thus gives $R_v = \frac{e}{2}\sqrt\frac{\gamma_1}{\gamma_2}$. In the limit where relaxation dominates coherence ($\gamma_1 = 2\gamma_2$), $R_v= \frac{e}{\sqrt{2}}\approx 1.922$.

We achieve a larger advantage, especially when $\gamma_1 < 2\gamma_2$, by using a state with a larger $v_x(0)$ and thus a finite breakdown time $t_b$. We stabilize $v_x(t)$ until breakdown and then set $h_y = 0$. In the small detuning limit, this gives improvement ratio
\begin{equation}
\label{eq:maxR}
R_v= e^{1-e^{-\gamma_2 t_b}}v_x(0) .
\end{equation}
While there is a closed-form expression for $t_b$ in terms of $v_x(0)$ derived in \cite{QPP}, it does not allow an analytical solution for the maximum of \cref{eq:maxR} over all values of $v_x(0)$. Instead we optimize numerically, as discussed below, reaching a maximum of $R_v = 1.96$ when $\gamma_1 = 2\gamma_2$. Even in the limit of no relaxation where $\gamma_1 \rightarrow 0$, we find $R_v= 1.09$. Thus, our protocol can achieve an unconditional $v_y$ signal boost compared to Ramsey interferometry.

When the inactive time $t_i$ is negligible and \cref{eq:ramsey-signal2} applies, we instead maximize the signal per root evolution time $v_y^c / \sqrt{t}$ over all times and initial states. In this case, using a permanently-stabilized state gives a maximum achievable advantage (when $\gamma_1 = 2 \gamma_2$) of only $R_s = 1.052$, and can be disadvantageous when dephasing is non-negligible. However, using an initial state with a finite breakdown time leads to a larger and unconditional SNR enhancement. Once again we can find an analytic expression for the maximum SNR and improvement ratio in terms of the initial state $v_x(0)$ and breakdown time $t_b$, but must numerically optimize over $v_x(0)$, as discussed in \cref{app:stable,app:unstable}. We find $R_s \geq 1$ for all values of $\gamma_1/\gamma_2$, reaching a maximum $R_s = 1.184$ when $\gamma_1 = 2 \gamma_2$.

\begin{figure}
    \centering
    \includegraphics[width=3.3in]{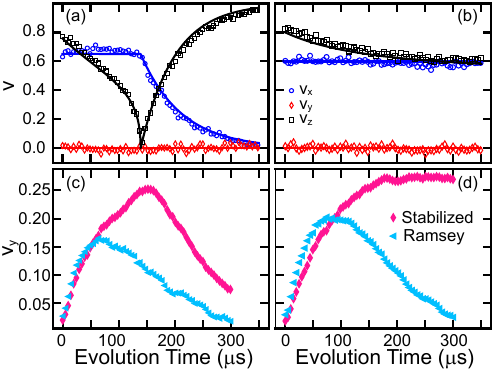}
    \caption{\label{fig:stabilized}
    (a-b) Bloch vector evolution, as measured through quantum state tomography, for a state with (a) $v_x(0) = 0.652$ and breakdown at $148 \ \mu$s and (b) $v_x(0) = 0.599$ with no breakdown (i.e, a stable state). (c-d) Evolution of $v_y$ for the same initial states as in (a-b), with added small detuning $\Delta/2\pi = 396$ Hz (c) and 324 Hz (d). Crucially, the stabilized states exhibit significantly larger signal than the Ramsey value.
    }
\end{figure}

\section{Experimental and Numerical Demonstration}\label{sec:exp}
We demonstrate our protocol using a superconducting qubit; device parameters and experimental details are given in \cref{app:expdetails}. We first show coherence stabilization with $\Delta = 0$. The experimental pulse sequence is shown in \cref{fig:cartoons}. We prepare a state in the $xz$ plane with $\vec{v}(0) = (\sin \theta,0,  \cos \theta)$. We then continuously drive the qubit to rotate the Bloch vector about the $y$ axis towards the $x$ axis. If the control exceeds maximum output amplitude of our control electronics (near breakdown), we set it to 0 for the remainder of the evolution. We cut off the control after breakdown to prevent rotations from $v_x$ to $v_z$ that would decrease $v_x$ faster and reduce the growth of our $v_y$ signal.

Quantum state tomography data showing coherence stabilization for two different initial states is shown in \cref{fig:stabilized}(a-b), along with theoretical predictions (not fits) generated by solving the Bloch dynamics for our system parameters. Depending on the initial state and the ratio $\gamma_1/\gamma_2$, the stabilization may exhibit a breakdown (panel (a)) or long-time stability (panel (b)). When the qubit is detuned from the drive frequency by some small detuning $\Delta \ll \gamma_2$, $v_y(t)$ grows to some maximum. We set $\Delta/2\pi = 396$ Hz and $324$ Hz for the same states as \cref{fig:stabilized}(a-b), respectively, and measure $v_y$ (\cref{fig:stabilized}(c-d)). We compare to $v_y$ from Ramsey sequences ($v_x(0) = 1$) for the same detunings. For both states there is an enhanced $v_y$ signal compared to Ramsey, validating the essential aspect of our protocol. Note that these data were taken on different days and $T_2$ drifted from $73 \ \mu$s (c) to $89 \ \mu$s (d), which accounts for the larger signal in (d) despite a smaller $\Delta$.

\begin{figure*}
    \centering
    \includegraphics[width=6.6in]{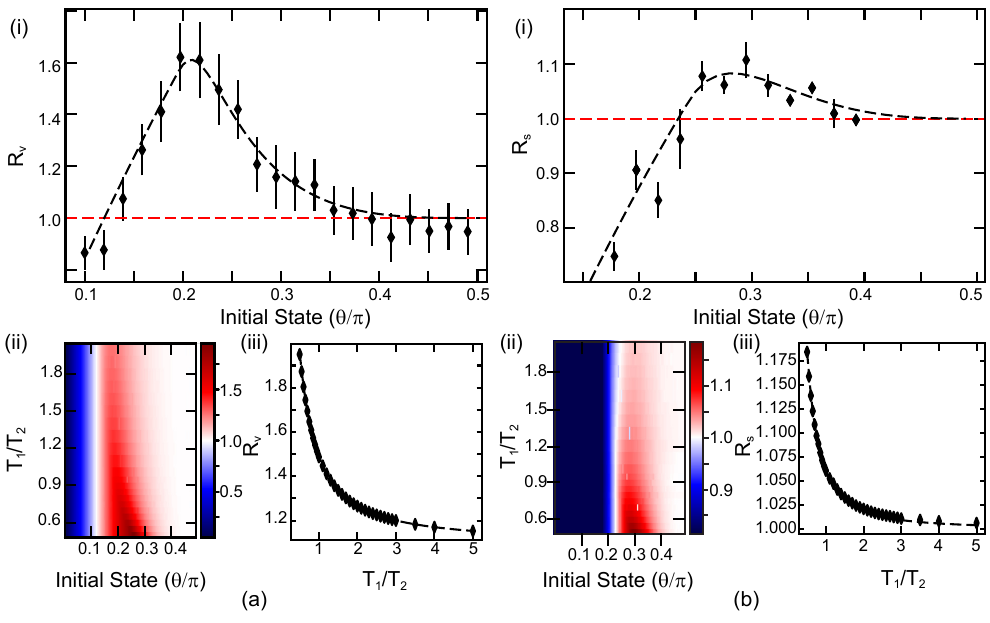}
 
    \caption{\label{fig:sweeps} Experimentally measured signal improvement ratio $R_v$ (a.i) and $R_s$ (b.i) as a function of initial state, measured at the times that theory predicts maximum signal for stabilized and Ramsey data. Error bars are calculated from the variance of the ratio across many measurements. The dashed line is a theoretical prediction, not a fit, showing good agreement. Bottom row: Numerically simulated improvement ratio $R_v$ (a.ii) and $R_s$ (b.ii) as a function of initial state and $T_1/T_2$ ratio, for small detuning. Also shown are numerically simulated (markers) and analytically derived (dashed line) improvement ratio $R_v$ (a.iii) and  $R_s$ (b.iii) at optimal initial state as a function of $T_1/T_2$.
     }
\end{figure*}

 To quantify the enhancement of signal $R_v$, we sweep the detuning $\Delta$ and measure the coherence-stabilized $v_y^c(t_{\max},\Delta,\theta)$  for each initial state polar angle $\theta$. Here, $t_{\max}$ is the predicted time of maximum $v_y^c$ (see \cref{app:stable,app:unstable}); for solutions with no breakdown we use $t_{\max} = 350 \ \mu\mathrm{s}\approx 5 T_2$. We also measure the Ramsey evolution $v_y^R(T_2,\Delta)$ at $t=T_2$ when theory predicts $v_y^R$ will be maximized. We then fit the slopes of $v_y^c$ and $v_y^R$ vs $\Delta$ and take their ratio to compute $R_v$. Results are plotted in \cref{fig:sweeps}(a.i). Using our protocol with initial state $\theta = 0.198\pi$ and $N$ shots, we are able to detect qubit frequency with a minimum 1-$\sigma$ uncertainty of $\sqrt{N}\delta f_c = 3.4 \pm 0.8 \ \mathrm{kHz} \sqrt{\mathrm{shots}}$, compared to $\sqrt{N}\delta f_\mathrm{R} = 5.5 \pm 0.7 \ \mathrm{kHz}\sqrt{\mathrm{shots}}$ for Ramsey (variances are over repetitions of the experiment; see \cref{app:SNR} for explanation of the sensitivity calculation and units). Thus our protocol reduces qubit frequency detection uncertainty by a factor of $R_v = 1.62 \pm 0.13$ when $t \ll t_i$. Theory predicts $v_y^c$ will be maximized when $\theta = 0.213\pi$ [$v_x(0) = 0.671$], with predicted $R_v = 1.649$. We find good agreement with the data with no free parameters, indicating that our protocol behaves as predicted.

To test the protocol under different environmental parameters, we numerically simulate the evolution as a function of initial state and $T_1/T_2$ ratio, all at small detuning $\Delta = 0.01 / T_2$. Results are plotted in \cref{fig:sweeps}(a.ii). We maximize over initial state at each $T_1/T_2$ and plot these maxima in \cref{fig:sweeps}(a.iii). We find that $R_v$ reaches a maximum value of 1.96 in the limit where relaxation dominates dephasing ($T_2 = 2T_1$), and has a minimum of 1.09 in the limit where dephasing dominates relaxation ($T_1 \gg T_2$), as predicted by analytical theory (shown as a dashed line). 

To quantify the enhancement of SNR per qubit evolution time $R_s$, we use a similar procedure as above, except that we measure the stabilized $v_y$ and Ramsey $v_y$ at the times theory predicts will maximize $v_y/\sqrt{t}$ (see \cref{app:stable,app:unstable}). Experimental, theoretical, and numerical results are plotted in \cref{fig:sweeps}(b). During these measurements, $T_1/T_2 = 0.764 \pm 0.156$ and $T_2 = 69.5 \pm 10.6 \ \mu$s. Using our protocol with initial state $\theta = 0.315 \pi$ and total experiment time $T$, we find a minimum frequency detection uncertainty of $\sqrt{T}\delta f_c = 63 \pm 4 \ \mathrm{Hz}/\sqrt{\mathrm{Hz}}$, compared to $\sqrt{T}\delta f_\mathrm{R} = 70 \pm 1 \ \mathrm{Hz}/\sqrt{\mathrm{Hz}}$ with Ramsey. Again, our protocol reduces uncertainty by a factor of $R_s = 1.11 \pm 0.32$ when $t_i \ll t$ (see \cref{app:SNR}). Our experimental results again agree well with the theory, which predicts a maximum $R_s=1.094$ at $\theta = 0.283\pi$ [$v_{x}(0) = 0.776$] for this $T_1/T_2$ ratio. Theory and simulation show the improvement ratio $R_s$ ranges from 1.184 when $T_2 = 2 T_1$ to 1 when $T_1 \gg T_2$.  

A quantum sensor may have comparable inactive time $t_i$ and evolution time $t$, between the limits we study. For a fixed $t_i$, the problem is to optimize $v_y(t)/\sqrt{t + t_i}$. Again, the Ramsey protocol can be optimized analytically, while our coherence stabilization protocol can be optimized numerically by the same procedure used above. The SNR improvement ratio will land somewhere between our two limiting cases, but will always be $\geq 1$.

\section{Protocol Robustness}

\begin{figure}
    \centering
    \includegraphics[width=3.3in]{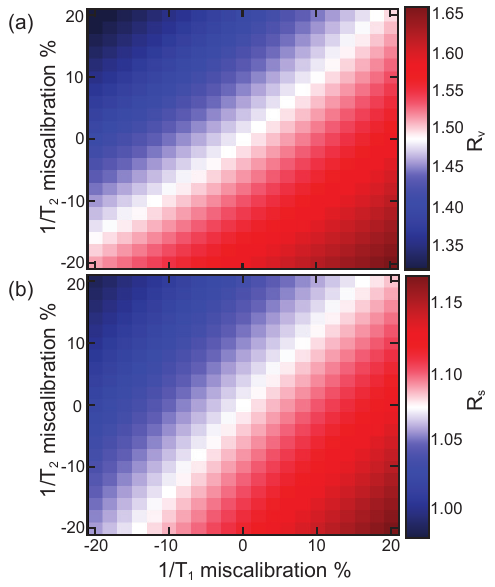}
    \caption{\label{fig:miscalibration}
    Simulated sensitivity of the improvement ratios (a) $R_v$ and (b) $R_s$ to miscalibrations of decoherence rates $1/T_1$ and $1/T_2$, for nominal $T_1/T_2 = 1$. Miscalibration of $1/T$ is defined as $(T_\mathrm{nominal}/T_\mathrm{actual} - 1)$. White color is defined as the value at the perfectly-calibrated center point.
    }
\end{figure}

Optimal sensing depends on accurate knowledge of $T_1$ and $T_2$. To quantify the robustness of our protocol to miscalibrations, we simulate the Bloch evolution using the initial state, control field, and measurement time that would be optimal for a nominal $T_1 = T_2 = 1$ while varying the actual values of $T_1$ and $T_2$ that control the dynamics. Thus, we simulate a situation in which $T_1$ and $T_2$ have changed unbeknownst to the experimenter. We simulate Ramsey experiments with the same miscalibration: measurements are performed at times determined by the nominal $T_2$, not the actual simulated one. We plot the SNR improvement ratios as a function of percentage change in $\gamma_1 = 1/T_1$ and $\gamma_2 = 1/T_2$ in \cref{fig:miscalibration}. We see little change in the improvement ratios when $T_1$ and $T_2$ are miscalibrated by the same factor, as shown by the near-constant values along lines of constant $T_1/T_2$ ratio running diagonally from bottom left to top right. We see some dependence of the improvement ratios on changes in the $T_1/T_2$ ratio, as shown by the steepest gradient running approximately diagonally from top left to bottom right. The majority of this dependence is not due to miscalibration, but rather is due to the fact that $R_v$ and $R_s$ depend on $T_1/T_2$ even when perfectly calibrated. This can be seen in the plots as the large regions where SNR is improved even more than at the perfectly calibrated $T_1$ and $T_2$ (red color), due to the fact that $T_1/T_2$ has decreased. There is some additional SNR suppression due to using a suboptimal initial state and measurement time, which is evident in the top left corner of \cref{fig:miscalibration}(b)---$R_s$ dips slightly below 1, indicating worse performance than Ramsey, while an optimal protocol would always have $R_s \geq 1$. Still, this requires quite a large deviation, with actual $T_2/T_1 \sim 2/3$ of its nominal value, and so the protocol is relatively robust to fluctuations in $T_1$ and $T_2$. Furthermore, we see from the experimental measurements in \cref{fig:sweeps} that improvement on par with the theoretical maximum value for that $T_1/T_2$ ratio can be achieved, even in a real system with fluctuating coherence times.

\section*{Conclusion}
In conclusion, we have demonstrated a protocol for enhancing qubit sensitivity to weak environmental fields by stabilizing partial qubit coherence. Our protocol requires only deterministic Hamiltonian control and is therefore applicable to a wide variety of qubit technologies. In the limit where decoherence is dominated by relaxation, we show a theoretical maximum of a  $1.96\times$ improvement over standard Ramsey interferometry in SNR per measurement shot, and a $1.184\times$ improvement in SNR per qubit evolution time. In our experimental apparatus with dephasing comparable to relaxation and with fluctuating system parameters, we achieve improvements of $1.6\times$ and $1.1\times$, respectively. Our results show a resource-efficient, broadly-applicable technique for unconditionally enhancing the SNR from qubit-based sensors and speeding calibration of qubit parameters.

While our technique provides a significant SNR boost over Ramsey interferometry, it is not fully optimal. For instance, if $\gamma_1 > 0$ and the optimal measurement time is after breakdown, there will be some nonzero $v_z$ that could be used to boost signal. Thus, for each set of environmental conditions, there is likely a Bloch trajectory (i.e., an initial state and control Hamiltonian) that provides an even larger signal than our protocol of stabilizing coherence. This is a problem of optimal control, and as such can be tackled with control theory techniques \cite{roloffOptimalControlOpen2009,remboldIntroductionQuantumOptimal2020}. It is a relatively unconstrained problem, as the initial state, final state, and final time are all variable. Given this lack of constraint, numerical solution methods will likely be required. Such optimal control has already been pursued in quantum sensing of time-varying signals \cite{poggialiOptimalControlOneQubit2018a,titumOptimalControlQuantum2021a}, and inspiration can be drawn from these results. In addition, it should be possible to stabilize properties of multi-qubit states~\cite{QPP}, including various entanglement measures~\cite{Horodecki:2009aa}.
Future work could therefore explore the possibility of extending our sensitivity enhancement to entangled states.

\acknowledgements{The authors gratefully acknowledge useful discussions with Kater Murch, Archana Kamal, Sacha Greenfield, and Sadman Shanto. Funding was provided by the National Science Foundation, the Quantum Leap Big Idea under Grant No. OMA-1936388, the Office of Naval Research under Grant No. N00014-21-1-2688, Research Corporation for Science Advancement under Cottrell Award 27550, and the ARO MURI grant W911NF-22-S-0007. Devices were fabricated and provided by the Superconducting Qubits at Lincoln Laboratory (SQUILL) Foundry at MIT Lincoln Laboratory, with funding from the Laboratory for Physical Sciences (LPS) Qubit Collaboratory.}

\bibliography{references}

\appendix
\renewcommand{\thefigure}{S\arabic{figure}}
\renewcommand{\thetable}{S\Roman{table}}
\stepcounter{myfigure}

\section{Experimental Details}\label{app:expdetails}

\begin{figure*}
    \centering
    \includegraphics[width=6.6in]{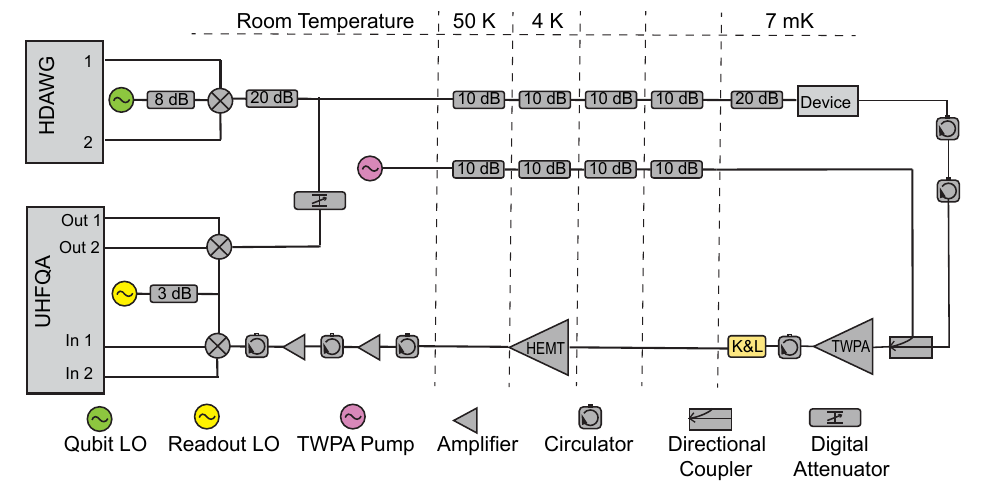}
    \caption{\label{fig:setup}
    Diagram of our cryogenic measurement setup. All qubit control envelopes are generated by the Zurich Instruments HDAWG, then upconverted and combined with measurement pulses from the ZI UHFQA before being fed into a heavily-attenuated line. Readout signals are amplified and then downconverted and fed back into the UHFQA for analysis.
    }
\end{figure*}

Our device is a standard grounded superconducting transmon qubit coupled to a quarter-wave transmission line cavity. Device parameters are given in \cref{tab:deviceparams}, and the design is available on the SQuADDS database \cite{shantoSQuADDSValidatedDesign2023}. The qubit and cavity are far off resonance. In this dispersive regime there is approximately 0 energy exchange between qubit and cavity, but the cavity frequency shifts by $\chi/2\pi = 150$ kHz when the qubit changes state. We measure the qubit by driving the cavity with an on-resonant pulse generated. The pulse transmits through and passes through an amplification chain up to room temperature, where we mix it back down to DC with an IQ mixer, giving a two-channel DC voltage signal that we then digitize. A diagram of the experimental setup is given in \cref{fig:setup}. We project the measured two-channel voltage onto an axis which gives maximum discrimination between the signals for qubit ground and excited states. We drive rotations of the qubit state by driving it with an on-resonance microwave pulse. The amplitude and duration of the pulse determine the total rotation angle, while the phase determines the rotation axis in the $xy$ plane.

We begin by measuring the qubit's $T_1$ (using a population decay measurement) and its $T_2$ and frequency (using a standard Ramsey measurement). If the qubit frequency has drifted, we reset the drive frequency to be resonant, then detune it by $\Delta$. We then calculate the control waveform to stabilize $v_x$ for our chosen initial state; in the case where we are stabilizing a state with a breakdown time and we want to extend the evolution past breakdown, we set the control to 0 after breakdown. We then calibrate the strength of our drives with a Rabi measurement, where we drive the qubit with a pulse of constant duration and variable amplitude. This pulse has a cosine envelope and is 2.35 $\mu s$ in duration. We measure oscillations of the qubit population and thus extract the driven Rabi frequency at a given control output voltage. We use this to convert our calculated control waveform into output voltage units for our control electronics. Note that long qubit manipulation pulses are used because our control line is heavily attenuated in order to give fine resolution of the continuous control waveform.

We use the same pulse envelope and duration for all qubit control pulses (except the continuous coherence-stabilizing drive)---we change the pulse rotation angle by changing the amplitude of the pulse, and change the rotation axis by changing the phase of the pulse. We therefore prepare a state with $v_x(0) = \sin \theta$ by driving a pulse with amplitude $\frac{\theta}{\pi} A_\pi$, where $A_\pi$ is the amplitude of a $\pi$ rotation pulse calibrated via the Rabi measurement. We next drive the continuous control waveform to stabilize the state for a time $t$. We then stop the control and perform quantum state tomography, measuring the qubit state along one axis. To measure $v_x$, we apply a $-\pi/2$ rotation pulse about the $y$ axis, then drive a readout pulse on the cavity. To measure $v_y$ we use a $\pi/2$ rotation about the $x$ axis, then a readout pulse; to measure $v_z$ we use a $2 \pi$ rotation about the $y$ axis, then readout. After a measurement we either do nothing or drive a $\pi$ rotation of the qubit, conditional on the measurement outcome, to reset it to the ground state. We wait an additional $60 \mu$s to damp any residual excited state population. We repeat each time point three times to measure all three Bloch vector components, then sweep $t$. Before measuring the first time point, we perform a measurement to calibrate the voltage corresponding to the ground state $v_z = 1$; after the last time point, we perform a $\pi$ rotation on the qubit and then measure to calibrate the voltage corresponding to the excited state $v_z = -1$. We then repeat this entire process many times and directly average the measured voltages. We use these voltages as the reference values for $v_i = \pm 1$ ($i = \{x,y,z\}$). 

When comparing the signal from coherence stabilization vs Ramsey, we interleave the measurements to avoid errors due to drifts in qubit parameters. We use threshold assignment of the readout voltage to the ground or excited state in order to reduce noise in these small signals. To speed data collection, we only measure $v_y$ and ignore the other Bloch components and only measure at the optimal times for maximizing $v_y$ and $v_y/\sqrt{t}$ ($T_2$ and $T_2/2$, respectively, for Ramsey, and the times derived below in \cref{eq:t'maxvy,eq:tmax,eq:t'max} for coherence-stabilized measurements). We interleave coherence-stabilized and Ramsey measurements for a given detuning and coherence-stabilized initial state, repeating many times to build up accurate estimates of $v_y$. We then move on to the next detuning while keeping $\theta$ constant. We repeat these detuning sweeps many times to build up statistics, re-measuring $T_1$ and $T_2$ before each sweep. We then move on to the next initial state $\theta$. 

We take this set of coherence-stabilized and Ramsey data from the many detuning sweep iterations for each $\theta$ and break it into chunks of $\sim 10$ iterations, interspersed throughout the dataset. For instance, one chunk might contain our 1st, 11th, 21st, ...,91st iteration. We rescale the detuning axis of this data by the $T_2$ measured in that iteration, and likewise divide the measurement times by $T_2$ to render them dimensionless. We then take all the $v_y$ data in a given chunk and fit it simultaneously to a linear dependence on $\Delta T_2$, extracting a slope for coherence stabilized and Ramsey measurements. In the case where we are calculating $R_s$, we then divide each slope by $t/T_2$, where $t$ is the time at which the measuremnent was taken in each iteration. Note that $t$ will vary linearly with $T_2$, so even as $T_2$ fluctuates through iterations, this ratio remains constant (so long as $T_1/T_2$ remains roughly constant). We take the ratio of the coherence-stabilized slope to the Ramsey slope, then average over all $N \sim 10$ chunks to give an estimate of $R_v$ or $R_s$. We compute the variance of these ratios and divide them by $N$ as an estimate of the error.

\begin{table}[]
    \centering
    \begin{tabular}{cccccc}
        $f_{01}$ & $\alpha / 2\pi$  & $f_\textrm{res}$ & $\kappa/2\pi$& Lamb Shift &$g/2\pi$\\
        (GHz)&(MHz) & (GHz) &(MHz) & (MHz) & (MHz)\\ \hline
        4.202 &-210 & 6.655&2.43 & 0.82 & 52 
    \end{tabular}
    \caption{Qubit Device Parameters}
    \label{tab:deviceparams}
\end{table}

\section{Bloch Equations Derivation}
\label{app:bloch}

Consider a single qubit with density operator $\rho$ subject to coherent evolution under the Hamiltonian $H$ and to environmental decoherence. Under the assumption that all decoherence is Markovian, we can model the dynamics with a Lindblad master equation:
\begin{equation}
    \dot{\rho} = -i\left[H,\rho\right]+\sum_\alpha \gamma_\alpha\left(L_\alpha \rho L_\alpha^{\dagger}-\frac{1}{2}\left\{L_\alpha^{\dagger} L_\alpha, \rho \right\}\right) ,
\end{equation}
where we take $\hbar = 1$. Let us consider the case where $H = h_y \sigma_y + \frac{\Delta}{2} \sigma_z$, the qubit has energy gap $\omega_{01}$ and is coupled to a thermal environment at inverse temperature $\beta$, 
the qubit undergoes dephasing at the rate 
$\gamma_\phi$
that can be modeled with the Lindblad operator $L_z = \sigma_z/\sqrt{2}$, relaxation at the rate 
$\gamma_r$
with corresponding Lindblad operator $L_1 = \sigma_-$, and excitation at the rate $e^{-\beta \omega_{01}}\gamma_r$ with corresponding Lindblad operator $L_2 = \sigma_+$. The factor $e^{-\beta \omega_{01}}$ reflects the Kubo-Martin-Schwinger condition (bath thermal equilibrium)~\cite{lidarLectureNotesTheory2020}. We relate the density operator to the Bloch vector $\vec{v}$ by
\begin{equation}
    \rho = \frac{1}{2}(I+\vec{v}\cdot\vec{\sigma}) ,
\end{equation}
where $I$ is the identity operator. Substituting in, 
and using $\Tr(\sigma_\alpha\sigma_\beta) = 2\delta_{\alpha\beta}I$, this leads to the Bloch equations
\begin{subequations}
\label{eq:vdotApp}
\begin{align}
   \label{eq:vxdotApp} \dot{v}_x &= -\gamma_2 v_x -\Delta v_y + 2h_y v_z \\
    \label{eq:vydotApp} \dot{v}_y &= -\gamma_2 v_y + \Delta v_x \\
    \label{eq:vzdottherm} \dot{v}_z &= \gamma_1(\eta-v_z) - 2h_y v_x
\end{align}
\end{subequations}
where 
\bes
\begin{align}
\gamma_1 &= (1+\exp(-\beta \omega_{01}))\gamma_r \\
\gamma_2 &= \gamma_\phi + \frac{\gamma_1}{2} ,
\end{align}
\ees
and where
\begin{equation}
\label{eq:eta}
\eta = \frac{1-\exp(-\beta \omega_{01})}{1+\exp(-\beta \omega_{01})} \in [0,1] ,
\end{equation}
with $\eta=1$ in the zero temperature ($\beta\to\infty$) limit.

In subsequent Appendix sections, we will use \cref{eq:vzdottherm} for modeling the finite temperature setup, where $\eta$ will carry the temperature dependence. Note that if $h_y = 0$ uniformly (e.g., in Ramsey experiments), then $v_x$ and $v_z$ are uncoupled, and therefore temperature does not affect $v_x$ given a fixed coherence dissipation rate $\gamma_2$.

The Bloch equations \eqref{eq:vdotApp} have a closed-form analytical solution for arbitrary detuning $\Delta$ and control $h_y$, but this solution is extremely complicated. The solution is drastically simpler when $\Delta=0$, since in this case $v_y(t) = v_y(0)e^{-\gamma_2 t}$, and we are left with two coupled equations for $v_x$ and $v_z$. When also $h_y=0$, we find $v_x(t) = v_x(0)e^{-\gamma_2 t}$ and $v_z(t) = \eta 
 + [v_z(0)-\eta]e^{-\gamma_1 t}$. For this reason, we define $T_1 \equiv 1/\gamma_1$ and $T_2 \equiv 1/\gamma_2$ as the relaxation and dephasing times. However, it should be noted that the decay times associated with $v_x(t)$ and $v_z(t)$ are  modified when $h_y\ne 0$. In this case, when $|\gamma_1-\gamma_2|>4|h_y|$ and $h_y$ is constant, we find both $v_x$ and $v_z$ include two decaying exponentials added together, with rates
\begin{align}
\gamma_{\pm a} &= \frac{1}{2} \left(\gamma_1+\gamma_2 \pm -\sqrt{(\gamma_1-\gamma_2)^2-(4h_y)^2}\right)
\end{align}
The amplitudes of each decaying component depend on $h_y$; when $h_y=0$, $v_x(t)$ retains only the $\gamma_{-a}\rightarrow\gamma_2$ term, and $v_z(t)$ retains only the $\gamma_{+a}\rightarrow\gamma_1$ term as expected. When $|\gamma_1-\gamma_2|\le 4|h_y|$ and $h_y$ is constant, both $v_x(t)$ and $v_z(t)$ oscillate at frequency $\sqrt{(4h_y)^2-(\gamma_1-\gamma_2)^2}/2$ and decay with the same rate of $(\gamma_1+\gamma_2)/2$. A time-dependent $h_y$ complicates the solution further, but always gives decay rates for $v_x$ and $v_z$ that depend on $\gamma_1$, $\gamma_2$, and $h_y$.

\section{Sensing Signal to Noise Ratio}\label{app:SNR}
We consider the quantum sensing problem of determining some environmental parameter $B$ to the best precision possible per total time $T$ spent performing the experiment. 
Consider a qubit with a resonantly tuned drive, subject to some environmental variable $B$ that shifts its frequency and thus causes detuning. That is to say, in the rotating frame of the drive where the qubit Hamiltonian $H_q = 0$, the system acquires a new Hamiltonian term $H'_q = -\frac{\Delta}{2} \sigma_z$. 
We write the initial qubit state in terms of its Bloch vector $\vec{v}(0) = (v_x(0),v_y(0),v_z(0))$ and set our coordinate system such that $v_y(0) = 0$. 
The effect of $H'_q$ is to rotate the Bloch vector about the $z$ axis at a rate $\Delta$.
For small rotation angle (i.e., for small $\Delta t$, where $t$ is the evolution time), $v_y$ grows linearly in $\Delta t$ proportionally to $v_x$. 
Measuring $v_y = \braket{\sigma_y}$---that is, measuring the probability $P(t) = \bra{+i}\rho(t)\ket{+i} = \frac{1}{2}[1+v_y(t)]$ to be found in the $\ket{+i}\equiv(\ket{0} + i\ket{1})/\sqrt{2}$ state given that the system is in state $\rho(t)$---
is thus a measurement of $\Delta$, since $t$ is set by the experimenter. Here, $v_y(0) = 0$ [$P(0) = 1/2$] and $v_y(\Delta t) = 2[P(\Delta t)-1/2]$. 
The signal is thus an estimate of the deviation of $v_y$ from its reference value $0$---equivalently, it is the deviation of the number of counts of the $\ket{+i}$ state from its reference value $N/2$, where $N$ is the number of iterations of the measurement. The SNR of this measurement is the expected value of excess $\ket{+i}$ counts divided by its standard deviation, 
\begin{align}\label{eq:SNR}
    \mathrm{SNR} =& \frac{CN(P-\frac{1}{2})}{\sqrt{NP(1-P)}} = \frac{CN\frac{v_y}{2}}{\sqrt{N\frac{(1+v_y)}{2}\frac{(1-v_y)}{2}}} \\
  &\approx C\sqrt{N}v_y, \notag
\end{align}  
where $C\leq 1$ accounts for state preparation and measurement errors that reduce the expected value, and the approximation is valid for $|v_y| \ll 1$ \cite{degenQuantumSensing2017}. In the main text we used $C=1$ under the assumption that $C$ would be the same for Ramsey and coherence-stabilized protocols, and would therefore divide off when we calculated the relative enhancement.

The fundamental uncertainty in a measurement of $v_y$ is $\frac{1}{C\sqrt{N}}$. We also have a linear relationship between the measured $v_y$ and $\Delta$ at a particular time, with $v_y = a \Delta$. We can thus equate the uncertainties 
\begin{align*}
    \delta {v_y} &= \frac{1}{C\sqrt{N}} = a \delta \Delta \longrightarrow \sqrt{N}\delta\Delta = \frac{1}{aC} \\
\end{align*}
We fit the slope $a$ in our data and thus extract the frequency detection uncertainty. In the case where number of shots is constant, we report an uncertainty in units of $\mathrm{Hz}\sqrt{\mathrm{shots}}$. One then divides by $\sqrt{N}$ to find the overall resolution in $\mathrm{Hz}$. In the case where idle time is negligible, we replace $N\rightarrow T/t$ and fill in $t$ as the evolution time from the experiment, while $T$ is set at the unit value of $1 \mathrm{s}$. We multiply both sides by $\sqrt{t}$ to find
\begin{align*}
    \sqrt{T} \delta \Delta &= \frac{\sqrt{t}}{C a}
\end{align*}
which gives an uncertainty in units of $\mathrm{Hz}/\sqrt{\mathrm{Hz}}$. For a given experiment time $T$ one then divides by $\sqrt{T}$ to find the overall resolution in $\mathrm{Hz}$. Note that $a$ depends on $T_2$, the protocol used, and the measurement time.

\section{Stable Solutions Improvement Derivation} 
\label{app:stable}

Here we present a full derivation of the control field, Bloch trajectory, and sensitivity enhancement using a state where coherence is stabilized indefinitely. We work in the rotating frame of the qubit drive, neglect higher transmon excited states and take the rotating wave approximation, which is generally valid for the weak control drives we apply (typical drive Rabi frequencies are $< 1$ MHz, while the qubit frequency is $\sim 4$ GHz). We restrict ourselves to the case where the qubit is initialized in the $xz$ plane and the control is about the $y$ axis, $H_\mathrm{drive} = h_y \sigma_y$. We also take the low-temperature limit where thermal qubit excitation is negligible. The dynamics of the Bloch vector are then given by the Bloch equations [\cref{eq:vdotApp}] with $\Delta$ representing the detuning as before.

In the long-time limit where the state saturates to a stable value, all derivatives in \cref{eq:vdotApp} are $0$ and the Bloch vector has an analytic solution, which we write in the zero temperature ($\eta=1$) limit:
\begin{subequations}
\begin{align}
    v_x &= \frac{2 \eta h_y \gamma_1 \gamma_2}{4 h_y^2 \gamma_2 + \gamma_1 (\gamma_2^2 + \Delta^2)} \\
    v_y &= \frac{2 \eta h_y \gamma_1 \Delta}{4 h_y^2 \gamma_2 + \gamma_1 (\gamma_2^2 + \Delta^2)} = \frac{\Delta}{\gamma_2}v_x\\
    v_z &= \frac{\eta \gamma_1 (\gamma_2^2 + \Delta^2)}{4 h_y^2 \gamma_2 + \gamma_1 (\gamma_2^2 + \Delta^2)}  
\end{align}
\end{subequations}
Maximizing $v_x$ thus maximizes $v_y$. In the limit of small detuning $\Delta\ll \gamma_2$, this occurs when $h_y = \sqrt{\gamma_1\gamma_2}/2$. Applying this control and solving for arbitrary $\Delta$ gives
\begin{equation}\label{eq:vymaxstab}
    v_y^{\max} = \frac{\sqrt{\gamma_1\gamma_2}\eta\Delta}{2\gamma_2^2+\Delta^2}
    \approx \frac{\eta}{2}\sqrt{\frac{\gamma_1}{\gamma_2}}\frac{\Delta}{\gamma_2}.
\end{equation}
On the other hand, if the state is initialized with $v_x = 1$ and allowed to decay freely (a Ramsey measurement), we have
\bes
\begin{align}
\label{eq:Ramsey-deriv}
    v_y^R(t) &= \sin({\Delta t})e^{-\gamma_2 t} \\
    v_y^{R,\max} &= \sin{\left(\tan^{-1}\left(\frac{\Delta}{\gamma_2}\right)\right)}e^{-\frac{\gamma_2}{\Delta}\tan^{-1}(\frac{\Delta}{\gamma_2})} \\
    &\approx \frac{1}{e}\frac{\Delta}{\gamma_2} ,
\end{align}
\ees
which is \cref{eq:ramsey-signal} in the main text.
This maximum value is achieved at time $\tan^{-1}(\frac{\Delta}{\gamma_2})/\Delta \approx T_2$. We see that in the limit of small detuning, 
\begin{equation}
    R_v\equiv \frac{{v_y}^{\max}_c}{{v_y}^{\max}_R} = \frac{\eta e}{2}\sqrt{\frac{\gamma_1}{\gamma_2}}.
\end{equation}
Thus, for a $T=0K$ (i.e., $\eta=1$) system, there is a signal boost in $v_y$ so long as $\frac{\gamma_1}{\gamma_2}>\frac{4}{e^2} \approx 0.54$.

In the case where measurement idle time is negligible, we instead wish to maximize $v_y(t)/\sqrt{t}$. In this case the Ramsey signal is maximized at $t = \frac{1}{2\gamma_2} = \frac{T_2}{2}$ with $v_y^R(t)/\sqrt{t} = (2 e)^{-1/2}$. To maximize the coherence-stabilized $v_y^c(t)/\sqrt{t}$ we solve for $v_y(t)$ assuming $v_x(t)=v_x(0)$:
\begin{equation}
 \label{eq:vystab}
     v_y^c(t) = (1-e^{-\gamma_2 t})v_x(0) \frac{\Delta}{\gamma_2}.
\end{equation}
We then take the largest stable value of $v_x$, divide \cref{eq:vystab} by $\sqrt{t}$, and maximize analytically. This maximum happens at 
\begin{equation}
\label{eq:tmax}
t_{\max} = -\frac{1}{2\gamma_2} \left(1 + 2 W_{-1}\left(-\frac{1}{2 \sqrt{e}}\right) \right) \approx 1.256 T_2
\end{equation}
where $W_k$ is the Lambert $W$ function, giving value
\begin{align*}\label{eq:stabSNR}
    \frac{v_y(t_{\max})}{\sqrt{t_{\max}}} &= \frac{\left(1 - e^{W_{-1}\left(-\frac{1}{2 \sqrt{e}}\right)+\frac{1}{2}}\right)}{\sqrt{- \left(2 W_{-1}\left(-\frac{1}{2 \sqrt{e}}\right)+1\right)}} \sqrt{\frac{\gamma_1}{2 \gamma_2}} \frac{\eta\Delta}{\sqrt{\gamma_2}} \\
    &\approx 0.319 \sqrt{\frac{\gamma_1}{\gamma_2}} \frac{\eta\Delta}{\sqrt{\gamma_2}}
\end{align*}
This leads to an SNR improvement ratio
\begin{equation}
    R_s \approx 0.742 \eta \sqrt{\frac{\gamma_1}{\gamma_2}}
\end{equation}
with maximum value $R_s = 1.052$ for $\eta=1$ when $\gamma_1 = 2 \gamma_2$, i.e., when dephasing is negligible. However, dephasing rates $\gamma_\phi > 0.05 \gamma_1$ cause $R_s$ to drop below 1, indicating no advantage. As shown below, using an unstable state yields a larger advantage that is always greater than 1.

\begin{figure}
    \centering
    \includegraphics[width=3.3in]{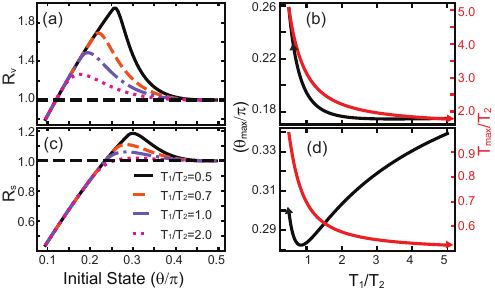}
    \caption{
    (a) Analytical theory curves for $R_v$ as a function of initial state at several $T_1/T_2$ ratios. (b) Numerical curves of initial state and measurement time at the values that maximize $R_v$, as a function of $T_1/T_2$ ratio. (c-d) Same as (a-b) but for $R_s$.
    }
    \label{fig:thycurves}
\end{figure}

\section{Unstable Solutions Improvement Derivation}\label{app:unstable}
In the case where we begin with a state that has a breakdown time, the dynamics are more complicated. We can simplify by treating $v_x$ as constant at $v_{x}(0)$ up until the breakdown time, which is valid in the limit of small detuning $\Delta$. In this case \cref{eq:vydotApp} is integrable and has solution \cref{eq:vystab}.
This equation holds until the breakdown time $t_b$ when $v_z=0$, at which time $v_y = v_y(t_b)$. After breakdown, we turn off the control, and so the state evolves freely according to \cref{eq:vxdotApp,eq:vydotApp} with $h_y=0$. This gives
\begin{equation}
    v_y(t_b + t^\prime) = [v_y(t_b) \cos(\Delta t') + v_x(0)\sin(\Delta t')] e^{-\gamma_2 t'}
\end{equation}
with $t' \equiv t-t_b$, i.e., the time since the breakdown. In the limit of small detuning, we can expand to first order in $\Delta$, giving
\begin{equation} \label{eq:afterbreak}
    v_y(t_b + t') = [v_y(t_b) + v_x(0) \Delta t']e^{-\gamma_2 t'}
\end{equation}
which is maximized at 
\begin{equation}\label{eq:t'maxvy}
    t'_{\max} = \frac{1}{\gamma_2} - \frac{v_y(t_b)}{\Delta v_x (0)} = \frac{1}{\gamma_2}e^{-\tau_b}
\end{equation}
where $\tau_b \equiv \gamma_2 t_b$ is the dimensionless breakdown time in units of $T_2$, which depends only on the initial state and the $T_1/T_2$ ratio. Thus $v_y$ reaches a maximum value 
\begin{equation}
    v_y^{\max} = e^{-e^{-\tau_b}}v_x(0) \frac{\Delta}{\gamma_2}.
\end{equation}
This gives an improvement ratio 
\begin{equation} \label{eq:imptbvx}
    R_v= e^{1-e^{-\tau_b}} v_x(0).
\end{equation}
All that remains is to maximize over the initial state $v_x(0)$. The breakdown time $t_b$ can be solved in terms of the initial state $\vec{v}(0)$ (assuming $v_z(0) > 0$):
\begin{align}
     t_b =& \frac{1}{2 \gamma_1} \left[\ln \left(\frac{\alpha^2+(2 v_z(0)-\eta)^2}{\alpha^2+\eta^2}\right) + \right. \notag \\
     & \left. \frac{2}{\alpha} \left(\tan ^{-1}\left(\frac{2 v_z(0)-\eta}{\alpha}\right)+\tan ^{-1}\left(\frac{\eta}{\alpha}\right)\right)\right] 
\end{align}
where we define
\begin{align}
     \alpha \equiv &\sqrt{\frac{4 v_x(0)^2 \gamma_2}{ \gamma_1} - \eta^2}
\end{align}
which quantifies the fractional difference between $v_x(0)$ and the largest stable value $v_x = \frac{\eta}{2}\sqrt{\frac{\gamma_1}{\gamma_2}}$. The breakdown time decreases with an increase in the temperature (i.e., $\eta$ decreases). This expression for $t_b$ does not allow for analytical maximization of \cref{eq:imptbvx}. Instead, we sweep the initial state and find the maximum value of $R_v$ over all initial states $v_x(0)$. Results are plotted in \cref{fig:thycurves}.

In the case where we consider negligible idle time in experiment, we instead divide \cref{eq:afterbreak} by $\sqrt{t}$ and then maximize. Here the maximum value occurs at a time $t_{\max} = t_b + t^\prime_{\max}$ with 
\begin{align}
\label{eq:t'max}
   t'_{\max} =  \bigg[ &\sqrt{4 e^{\tau_b}\left(2 \tau_b+1\right)+e^{2 \tau_b}\left(4 \tau_b\left(\tau_b+1\right)-7\right)+4} + \notag \\ 
  & 2 -e^{\tau_b}\left(2 \tau_b+1\right) \bigg] \frac{e^{-\tau_b }}{4}T_2.
\end{align}

Since the evolution after breakdown is decoupled between $v_x$ and $v_z$, $t^\prime$ is independent of $\eta$ (independent of temperature), except via the dependence on the breakdown time. Once again, this expression does not allow for analytical maximization of $v_y(t_b + t'_{\max})/\sqrt{t_b + t'_{\max}}$ over initial states and breakdown times, so instead we sweep the initial state $v_x(0)$ and numerically maximize. Results for the optimal initial state, optimal measurement time, and SNR improvement ratio $R_s$ (all as a function of $T_1/T_2$ ratio in the 0-temperature limit) are plotted in \cref{fig:thycurves}.

\end{document}